\documentclass{ws-procs975x65}            
\begin{document}
\title{Exploring gravitational statistics not based on quantum dynamical assumptions}

\author{P. A. Mandrin}

\address{Department of Physics, University of Zurich, 8057 
Z\"urich, Switzerland\\
$^*$E-mail: pierre.mandrin@uzh.ch}

\begin{abstract}
Despite considerable progress in several approaches to quantum gravity, there remain uncertainties on the conceptual level. One issue concerns the different roles played by space and time in the canonical quantum formalism. This issue occurs because the Hamilton-Jacobi dynamics is being quantised. The question then arises whether additional physically relevant states could exist which cannot be represented in the canonical form or as a partition function. For this reason, the author has explored a statistical approach (NDA) which is not based on quantum dynamical assumptions and does not require space-time splitting boundary conditions either. For dimension 3+1 and under thermal equilibrium, NDA simplifies to a path integral model. However, the general case of NDA cannot be written as a partition function. As a test of NDA, one recovers general relativity at low curvature and quantum field theory in the flat space-time approximation. Related paper: arxiv:1505.03719.
\end{abstract}


\bodymatter

\section{Introduction}
\label{sec:intro}

Although several approaches to quantum gravity have made considerable progress during the last decades, some fundamental issues are still under active consideration \cite{Carlip1}. One intriguing issue is that the quantisation procedure forces space and time to play different roles, in contrast to general relativity (GR). Following the idea of GR, one could think of possible physical states which are incompatible with space-time splitting and therefore cannot be represented by canonical quantisation or by an equivalent path integral description or by a statistical approach governed by a dynamical microscopic model (e.g. the quasi-local approach \cite{Brown_York_1992, Brown_York_1993, Creighton_Mann}). Such states would have to be constructed either by generalising the Hamiltonian formalism (introducing polymomenta \cite{Kanatchikov}) or, in the present study, by considering statistics not based on quantum dynamics (or NDA for short). The procedure is to compute the probability for every macroscopic state to happen under given constraints, to introduce a parameterisation of macroscopic states and then to vary the probability functional around its maximum. This procedure is also motivated by the related concept of black hole thermodynamics (Bekenstein \cite{Bekenstein} and Hawking \cite{Hawking}), and by the analogy between the second law of thermodynamics and Hamilton's variation principle.

This text summarises the method and first results and puts them into relation with other models. The interested reader will find a more complete introduction (and derivation) to this method \cite{Mandrin_NDA} which is self-contained and can be followed safely step by step. Former attempts to formulate a similar non-dynamical model exist \cite{Mandrin1, Mandrin4} but are not based on the same derivation and are not required for the reader to understand the present text. NDA is different from any former approaches which either assume quantum dynamics of some kind or else have no quantum model description at all and thus no straight-forward formulation.

\section{Partitioning of primary quanta}
\label{sec:part}

Consider a system $\mathcal{S}$ (a set) containing a statistically large number $N \gg \sqrt{N}$ of objects called primary quanta. To inspect part of $\mathcal{S}$, we partition $\mathcal{S}$ into $n$ subsystems $\mathcal{S}_i$ containing $N_i$ quanta, $i = 1 \ldots n$. Together with an arbitrary ordering (with relations $\mathcal{S}_j <: \mathcal{S}_k$), we obtain an (ordered) partitioning $\mathcal{P} = (\{\mathcal{S}_i\}; <:)$.

Define coarse-grained partitions $\mathcal{P}$ with $n$ subsystems and $N_j\gg \sqrt{N}_j$, $j=1\ldots n$, such that each $\mathcal{S}_j$ contains subsystems $\mathcal{S}^b_{j1} \ldots \mathcal{S}^b_{jq_j}$ as members of very fine partitions $\mathcal{P}^b$  with $q_j=n^b/n \ge 2$ and $n^b$ fixed, and call the $\mathcal{S}^b_{jl}$ boxes. Then, the partition $\mathcal{P}$ with the largest number $\Omega$ of possible fine partitions $\mathcal{P}^b$ is the one with the highest probability to occur, while other partitions are suppressed. We choose $\mathcal{P}$ fine enough in order to resolve any relevant constraints on the subsystems of $\mathcal{S}$ (e.g. due to observational data). To make $\Omega$ finite, we must impose a constant $p-1 \ge 1$ but not much greater than 1 before the computations and restrict the population of each box by setting $N^b_i \le p-1$. We may define the ''temperature'' $T_j$ of each $\mathcal{S}_j$ as follows. We insert into $\mathcal{S}_j$ a few additional boxes filled by a few quanta, thus varying $N_j$ and the entropy $S_j = \ln{\Omega}_j$ of $\mathcal{S}_j$. Then we define $T_j = \frac{\delta E_j}{\delta S_j}$, where $E_j = \ln{p} \cdot N_j$. The entropy $S_j$ trivially satisfies

\begin{equation}
\label{eq:entropy}
S_j = n_j \ \ln{p}; \qquad \delta S_j = \frac{\delta E_j}{T_j}.
\end{equation}

\noindent  An unconstrained system is in thermal equilibrium if the temperatures of all its statistically large subsystems are equal. However, if constraints are imposed on the subsystems, the system generally fails to be in thermal equilibrium, even for maximum entropy, because the constraints remove part of the possible macroscopic states of the system.

We now parameterise the coarse-grained subsystems $\mathcal{S}_j$ by assigning to each one a set of numbers $(x_j^k)$. It is convenient to increment the $k$th component $x_j^k$ to obtain the $k$th covering of $\mathcal{S}_j$. If $\mathcal{S}_j$ has $m_j$ coverings, $(x_j^k)$ can be seen as an element of a local vector space $V_j$ isomorphic to $R^{m_j}$. For the partition $\mathcal{P}$ with the highest probability, it can be shown that all values $m_j$ must be equal, $m_j = d =$ constant, and $d$ is called the dimension. One can also show that every path along arbitrary coverings is closed after a finite sequence of coverings.


\section{The second law of thermodynamics}
\label{sec:second}

The partition with the highest probability corresponds to the macroscopic state we mostly expect to observe. Thus, we maximise the entropy $S= \sum_j S_j$ of $\mathcal{S}$ under constraints, using Lagrange multipliers. Because $\mathcal{P}$ is chosen as fine as possible and every path can be closed, the sum can be converted to a closed integral over the domain $\mathcal{T}  = x(\mathcal{S})$ and then maximised: 

\begin{equation}
\label{eq:S_Int}
\delta S_c = \delta \oint_\mathcal{T} {\rm d}^d x \ [s(x^k) + \sum_{l=1}^{m_c} \ \lambda_l(x^k) \ \zeta_l(x^k)] = 0
\end{equation}

\noindent with entropy density $s(x^k)$, $m_c$ constraint functions $\zeta_l(x^k)$ and Lagrange multiplier functions $\lambda_l(x^k)$. The closed integral may be interpreted as a boundary integral, and there must be a periodic identification of points which we  are free to encode by substituting ${\rm d}^dx = {\rm  i}{\rm d}^dx_L$ in (\ref{eq:S_Int}) (Wick rotation). $x^k=x^k_E$ are Euclidean, $x^k_L$ Lorentzian-like parameters (we shall often ommit the subscripts).

We shall consider here smooth functions $s(x^k)$, $\lambda_l(x^k)\zeta_l(x^k)$, so that the tools of differential geometry apply. We are free to replace the parameterisation via an invertible transformation $x^k \rightarrow {x'}^k$, which is smooth and thus a diffeomorphism.

We next apply Gauss' law to (\ref{eq:S_Int}). If the boundary is non-orientable, it can be replaced by an orientable boundary without changing $S_c$, and Gauss' law can be extended. Via diffeomorphisms, we define the $d$-dimensional vielbeins $e^L_l$, the metric $\gamma_{kl}$, the ($d+1$)-dimensional vielbeins $e^\Gamma_\gamma$ and the metric $g_{\mu\nu}$ \cite{Mandrin_NDA}. Due to the freedom of parameterisation, we also vary with respect to the vielbeins in (\ref{eq:S_Int}). It then becomes convenient to define $\xi_l(x^k) = \gamma^{-1/2}\zeta_l(x^k)$, $\tau^{k\gamma}_K n_\gamma = \gamma^{-1/2} \ s \ \gamma^{kl}e^L_l \ \eta_{LK}$ with unit vector $n_\gamma$ normal to the space $V_j$  of $\mathcal{S}_j(x^k)$ and $\kappa^\mu n_\mu = \sum_{l=1}^{m_c} \ \lambda_l \ \xi_l$. We thus obtain

\begin{eqnarray}
\delta \oint_\mathcal{T} {\rm d}^d x \sqrt{\gamma} \ [\gamma^{-1/2} \ s + \sum_{l=1}^{m_c} \ \lambda_l \ \xi_l] & = & 0, \nonumber \\
\label{eq:S_vielbein}
\delta \oint_\mathcal{T} {\rm d}^d x \ n_\gamma \sqrt{\gamma} \ [\tau^{k\gamma}_K \ e^K_k + \kappa^\gamma] & = & 0. 
\end{eqnarray}

After applying Gauss' theorem, one obtains a field equation of the form of a higher curvature generalisation of GR (Palatini notation), in Lorentz-like coordinates:

\begin{equation}
\label{eq:motion}
\delta \int_\mathcal{M} {\rm d}^{d+1} x \ \sqrt{-g} \ [e^\Delta_\mu e^\Gamma_\nu \Phi^{\mu\nu}_{\Delta\Gamma} + \mu] = 0.
\end{equation}

\noindent with $\Phi^{\mu\nu}_{\Delta\Gamma} = \eta_{\Delta\Gamma} g^{\nu\delta} \tilde{\nabla}_\gamma [e^\Lambda_\delta \tau^{\mu\gamma}_\Lambda]$, $\partial \mathcal{M} = \mathcal{T}$, $\mu = \tilde{\nabla}_\mu \kappa^\mu$, and $\tilde{\nabla}_\gamma$ is the torsionless covariant derivative. (\ref{eq:motion}) can be shown to reduce to GR under weak field conditions, for negligible torsion, $d=3$ and introducing the known constants of GR (Newton's and cosmological constant) \cite{Mandrin_NDA}, in accord to available experimental data. Here, we briefly outline how this can be proven.

We first introduce the quantity 

\begin{equation}
\rho = \nabla_\gamma(\tau^{\delta\gamma}_\Delta \ e^\Delta_\delta ) = \nabla_\gamma(\tau^{\alpha\beta\gamma} \ g_{\alpha\beta}) = \rho^{\alpha\beta} \ g_{\alpha\beta}
\end{equation}

\noindent with $\tau^{\alpha\beta\gamma} = \tau^{\alpha\gamma}_\Gamma \ \eta^{\Gamma\Delta} \ e_\Delta^\beta$, $\nabla_\gamma g_{\alpha\beta} = 0$ and $\rho^{\alpha\beta} = \nabla_\gamma\tau^{\alpha\beta\gamma}$.

\noindent Using the equation

\begin{equation}
 \delta \int_\mathcal{M} {\rm d}^{d+1} x \ \sqrt{-g} \ [\rho + \mu] = 0,
\end{equation}

\noindent it can be shown by standard manipulations that, up to a divergence form,

\begin{equation}
\int_\mathcal{M} {\rm d}^{d+1} x \ \sqrt{-g} \ [\chi_{\mu\nu} + \theta_{\mu\nu}] \ \delta g^{\mu\nu} = 0,
\end{equation}

\noindent where $\chi_{\mu\nu} = \rho_{\mu\nu} - \rho g_{\mu\nu} / 2$ and $\theta^{\mu\nu} = \frac{2}{\sqrt{-g}} \ \frac{\delta (\sqrt{-g} \mu)}{\delta g_{\mu\nu}}$. By performing an infinitesimal translation $x^\mu \rightarrow x^\mu + \epsilon a^\mu(x^\nu)$ and integrating $\theta_{\mu\nu}$ over a region $\mathcal{M}$ with $\theta^\lambda_\nu a^\nu = 0$ on the boundary, one can show that $\theta^{\mu\nu}$ and $\chi^{\mu\nu}$ are divergence-free. Furthermore, considering contributions to $\chi_{\mu\nu}$ no higher than quadratic in the dimension of derivatives of $g_{\mu\nu}$, one can show that $\chi_{\mu\nu}$ only may have terms $\sim G_{\mu\nu}$ and $\sim g_{\mu\nu}$, where $G_{\mu\nu}$ is the Einstein tensor and the second contribution has the form of the cosmological constant term. This concludes the outline of the proof.

By analogy to GR, the quantity $\mu$ can be interpreted as the ''matter term'' (corresponding to $T^\mu_\mu$). Any symmetries imposed to the parameterisation contribute some part $\mu_{\rm psc}$ to $\mu$, and $\mu_{\rm psc}$ satisfies these symmetries. This is similar to the symmetry properties of the QFT Lagrangian of a free matter field on a background space and suggests us to define dual fields $\psi_l(x^k)$ so that $\mu_{\rm psc} \sim \sum_l |\psi_l|^2$.


\section{Quantum formalism}
\label{sec:quantum}

Experimental evidence of quantum behaviour involves detectors which are sensitive to a narrow selection out of $n_D$ possible quantum processes $P1 \ldots Pn_D$. Each detector represents a macroscopic system of many quanta. In order to predict the occurrence of e.g. $P1$, we need to compute the probability of detecting $P1$ rather than $P2 \ldots Pn_D$. For any channel $Pk$, $k=1\ldots n_D$, we determine the macroscopic state with the largest number $\Omega_{Pk}$ of microstates under the observational data constraint $\mu_{Pk}$ by evaluating $\delta S_c\big|_{\mu=\mu_{\rm psc}+\mu_{Pk}} = 0$. Then, the probability for $P1$ is

\begin{equation}
\label{eq:prob} 
p(P1) = \frac{\Omega_{P1}}{\sum_{k=1}^{n_D}\Omega_{Pk}}.
\end{equation}

\noindent This procedure differs from the path integral method. To obtain the path integral, consider the approximation of local thermal equilibrium as given by the partition function $Z(V,T)$ (canonical) for a sufficiently small ($d+1$)-volume $V$:

\begin{equation}
\label{eq:Z_e}
Z(V, T) \approx \int [\prod_{\Gamma,\gamma} {\rm d} e^\Gamma_\gamma] \ [\prod_{\mu,\Delta\le\Lambda} {\rm d} \omega_{\mu\Delta\Lambda}] \ {\rm e}^{{\rm i} S_c(V,T)},
\end{equation}

\noindent where $\omega_{\mu\Delta\Lambda}$ is the connection one-form. Integrating $Z(V, T)$ over a large volume yields an expression of the form ($\tilde{e}^\Gamma_\gamma$, $\tilde{\omega}_{\mu\Delta\Lambda}$ equal $e^\Gamma_\gamma$, $\omega_{\mu\Delta\Lambda}$ times scaling factors)

\begin{equation}
\label{eq:pi} 
Z_{\rm tot} = \int [\prod_{\Gamma,\gamma} \mathcal{D} \tilde{e}^\Gamma_\gamma] \ [\prod_{\mu,\Delta\le\Lambda} \mathcal{D} \tilde{\omega}_{\mu\Delta\Lambda}] \ {\rm e}^{{\rm i} S_c(V,T)}.
\end{equation}

\noindent Ignoring the connection part, one recovers the path integral formulation of torsionless gravity. This formulation applies to a slowly varying mean curvature, but not in the case of locally unbounded gradients of the curvature and therefore of $s$.

In the special case of the flat ($d+1$)-parameter space approximation, the approximation of local thermal equilibrium applies trivially and yields an expression of the form of the integral path representation of quantum field theory \cite{Mandrin_NDA} (for $\mu = \mu_{\rm psc}$):

\begin{equation}
\label{eq:interm}
Z(V_{\rm tot}) \approx \lim_{K\rightarrow \infty} \prod_{k=1}^K Z(V_k,T_k) = \int [\prod_l \mathcal{D}\psi_l] \ \exp({\rm i}\int_{V_{\rm tot}} {\rm d}^{d+1}x \ \mu_{\rm psc}).
\end{equation}

\noindent QFT corresponds to the case $d=3$ and $\hbar \mu_{\rm psc}$ corresponds to the Lagrangian of the matter fields. By inspection, we can identify $\hbar$ as the amount of parameterisation space required by one box if we set $p-1=1$. From (\ref{eq:motion}), we also find: The more matter $\mu_{\rm psc}$ we have per portion of space, the higher the curvature will be. We may thus conjecture that the maximum of $E$ per box is related to the Planck energy.


\bibliographystyle{ws-procs975x65}
\bibliography{ndamg14}

\end{document}